\documentclass{appolb}

\usepackage{amssymb}
\usepackage{amsmath}
\usepackage{epsfig}
\usepackage{amsfonts}
\usepackage{graphicx}%

\begin{document}
\title{Phenomenology of Dilaton in a Chiral Linear Sigma Model with Vector Mesons%
\thanks{Presented at the Workshop \textquotedblleft Excited QCD 2012\textquotedblright,
Peniche/Portugal, 6-12 May 2012 }%
}
\author{Stanislaus Janowski
\address{Institute for Theoretical Physics, Goethe-University,
Max-von-Laue-Str.\ 1, D--60438 Frankfurt am Main, Germany}
}
\maketitle
\begin{abstract}
In the framework of the $U(2)_{R}\times U(2)_{L}$ symmetric linear sigma model
with (axial)vector mesons generalized by including a dilaton field we study
the phenomenology of the scalar-isoscalar resonances below $2$
GeV. It turns out that, in our favoured scenario, the resonance $f_{0}(1370)$ is
predominantly a $\bar{q}q$ state and $f_{0}(1500)$ is predominantly a glueball
state. Additionally we
are able to calculate the value of the gluon condensate, which is in agreement
with lattice QCD results.
\end{abstract}

\section{Introduction}

One of the interesting issues in particle physics is the overpopulation in the
scalar-isoscalar channel, $I^{G}(J^{PC})=0^{+}(0^{++})$ in the energy region below
$2$ GeV. In this region there are currently five states listed by PDG \cite{PDG}.
Below $1$ GeV there are two resonances: $f_{0}(500)$ and $f_{0}(980)$.
Above $1$ GeV there are the three resonances $f_{0}(1370),f_{0}(1500)$ and $f_{0}(1710)$.
The nature of these resonances is not completely understood up to the present day.
In this work we consider an $N_{f}=2$ \cite{Paper1, dilaton} effective Lagrangian
with two scalar-isoscalar states, where one of them is a quark-antiquark state,
$\left\vert \bar{n}n\right\rangle \equiv\left(\bar{u}u+\bar{d}d\right)  /\sqrt{2},$
and the other one is a scalar glueball, $\left\vert G\right\rangle \equiv gg.$
The experimental verification of glueballs is extremely challenging due to the
mixing with ordinary $\bar{q}q$ mesons, but the proof of their existence would be a further
important achievement in QCD. Furthermore, the proper candidate for a glueball
should possesses the following features: i) Due to the `democratic' coupling
of the gluons to all kinds of quarks, the glueball should be flavour blind.
ii) The decay width should be rather narrow because the large-$N_{c}$ behaviour
shows that glueball decays scales as $N_{c}^{-2}$, thus they are stronger suppressed
than decays of ordinary mesonic $\bar{q}q$ states, which only scale as $N_{c}^{-1}.$
The flavour blind decay behaviour of the resonance $f_{0}(1500)$ with a mass of
$M_{f_{0}(1500)}=(1505\pm6)$ MeV and its narrow decay width,
$\Gamma_{f_{0}(1500)}=\left(  109\pm7\right)$ MeV \cite{PDG}, make it a proper
scalar glueball candidate. The main aim of this study is to
investigate the mixing between the scalar-isoscalar glueball and the scalar-isoscalar
quark-antiquark state in order to make some statements about the nature of the
scalars-isoscalars below the mass of $2$ GeV.\newline
As shown in Ref \cite{dilaton} on which
this proceeding is based on, our most successful scenario is realized
by the assignment: $\left\vert G\right\rangle $ $\cong f_{0}(1500)$ and $\left\vert \bar
{q}q\right\rangle $ $\cong f_{0}(1370)$. Moreover, due to the new available determination
of the resonance $f_{0}(500)$ \cite{PDG} ($M_{f_{0}(500)}=(400-550)$ MeV
and $\Gamma_{f_{0}(500)}=(400-700)$ MeV) we tested again the following two
alternative scenarios: 1) $\left\vert G\right\rangle $ $\cong f_{0}(1500)$ and
$\left\vert \bar{q}q\right\rangle $ $\cong f_{0}(500)$ and 2) $\left\vert
G\right\rangle $ $\cong f_{0}(1710)$ and $\left\vert \bar{q}q\right\rangle $
$\cong f_{0}(500),$ but both of them are still inconsistent with the experimental
data and hence not favoured. Additionally, by the use of the phenomenology of
mesons only, we are capable to calculate the value of the gluon condensate,
$\left\langle \frac{\alpha_{s}}{\pi}\,G_{\mu\nu}^{a}G^{a,\mu\nu}\right\rangle $,
which agrees to the lattice QCD results. 

\section{The Model}

The effective model under study is based on the chiral symmetry\linebreak
$U(N_{f})_{R}\times U(N_{f})_{L}$ and on the trace anomaly of QCD. It is
composed of the quark-antiquark Lagrangian (see Ref. \cite{Paper1}) and of the dilaton
Lagrangian describing the trace anomaly (see Ref. \cite{dilaton} and
refs. therein). The latter one reads:%
\begin{equation}
\mathcal{L}_{dil}=\frac{1}{2}(\partial_{\mu}G)^{2}-\frac{1}{4}\frac{m_{G}^{2}%
}{\Lambda^{2}}\left(  G^{4}\ln\left\vert \frac{G}{\Lambda}\right\vert
-\frac{G^{4}}{4}\right)  \text{ },\label{ldil}%
\end{equation}
where $G$ is the scalar dilaton/glueball field \cite{schechter} and $\Lambda$
an energy scale. The minimum of the potential in Eq. (\ref{ldil}) is given by
$G_{0}=\Lambda$, and after shifting the dilaton field, $G\rightarrow G_{0}+G,$ a
massive particle occurs, which corresponds to the scalar glueball. According to
lattice QCD \cite{Morningstar} its mass is about $m_{G}\sim1.6$ GeV.
The logarithmic term breaks the dilatation symmetry,
$x^{\mu}\rightarrow\lambda^{-1}x^{\mu}$, explicitly and this leads to the
divergence of the corresponding current: $\partial_{\mu}J_{dil}^{\mu
}=T_{dil,\,\mu}^{\;\mu}=-\frac{1}{4}m_{G}^{2}\Lambda^{2}.$ In the chiral limit
(neglecting the $U(1)_{A}$ anomaly) the energy scale $\Lambda$ in Eq. (\ref{ldil})
is the only dimensionful parameter of the effective model. Furthermore, we require
that the effective Lagrangian must be finite for every finite value of the gluon
condensate $G_{0}$. This implies that in the chiral limit all other terms have dimension
$\left[\text{E}^{4}\right]  $ in order to ensure dilatation invariance. Thus, the full
effective Lagrangian reads:%
\begin{align}%
\mathcal{L}%
&  =\mathcal{L}_{dil}+\text{\textrm{Tr}}\left[  (D^{\mu}\Phi)^{\dag}(D_{\mu
}\Phi)-m_{0}^{2}\left(  \frac{G}{G_{0}}\right)  ^{2}\Phi^{\dag}\Phi
-\lambda_{2}(\Phi^{\dag}\Phi)^{2}\right]  -\lambda_{1}([\text{\textrm{Tr}}%
\Phi^{\dag}\Phi])^{2}\nonumber\\
&  +c_{1}[\text{\textrm{det}}(\Phi)-\text{\textrm{det}}(\Phi^{\dag}%
)]^{2}+\mathrm{Tr}[H(\Phi^{\dag}+\Phi)]-\frac{1}{4}\text{\textrm{Tr}}\left[
(L^{\mu\nu})^{2}+(R^{\mu\nu})^{2}\right]  \nonumber\\
&  +\frac{m_{1}^{2}}{2}\left(  \frac{G}{G_{0}}\right)  ^{2}\text{\textrm{Tr}%
}\left[  (L^{\mu})^{2}+(R^{\mu})\right]  ^{2}+\frac{h_{1}}{2}\text{\textrm{Tr}%
}[\Phi^{\dag}\Phi]\text{\textrm{Tr}}[L_{\mu}L^{\mu}+R_{\mu}R^{\mu}]\nonumber\\
&  +h_{2}\text{\textrm{Tr}}[\Phi^{\dag}L_{\mu}L^{\mu}\Phi+\Phi R_{\mu}R^{\mu
}\Phi^{\dag}]+2h_{3}\text{\textrm{Tr}}[\Phi R_{\mu}\Phi^{\dag}L^{\mu}]\text{
,}\label{Lagrangian}%
\end{align}
where the (pseudo)scalar and the left- and the right-handed (axial)vector
d.o.f. are organized in multiplets. The explicit form of these multiplets in
the case of $N_{f}=2$ is the following: $\Phi=(\sigma+i\eta_{N})\,t^{0}%
+(\vec{a}_{0}+i\vec{\pi})\cdot\vec{t}$ , $L^{\mu}=(\omega^{\mu}+f_{1}^{\mu
})\,t^{0}+(\vec{\rho}^{\mu}+\vec{a}_{1}^{\mu})\cdot\vec{t}$ and $R^{\mu
}=(\omega^{\mu}-f_{1}^{\mu})\,t^{0}+(\vec{\rho}^{\mu}-\vec{a}_{1}^{\mu}%
)\cdot\vec{t}$ , where $t^{0}$, $\vec{t}$ are the corresponding generators.
The assignment of the fields in (\ref{Lagrangian}) is as follows: The
pseudoscalar fields $\vec{\pi}$ and $\eta_{N}\equiv(\overline{u}u+\overline
{d}d)/\sqrt{2}$ with $m_{\eta_{N}}=716$ MeV \cite{Paper1,Giacosa:2007up}
represent the pions \cite{PDG} and the nonstrange part of the
$\eta$ meson, respectively. The scalar-isoscalar field, $\sigma\equiv\left(
\bar{u}u+\bar{d}d\right)  /\sqrt{2}$, represents the nonstrange quark-antiquark
state. It turns out that the resonance $f_{0}(1370)$ is favoured to be a
predominantly $\bar{q}q$ state \cite{Paper1}. Therefore we identify $\sigma$
with the resonance $f_{0}(1370)$, but we also assign it to the resonance
$f_{0}(500)$ in order to test all possible scenarios. Corresponding to the
study of \ Ref. \cite{Paper1} the scalar-isovector fields $\vec{a}_{0}$
represent the resonance $a_{0}(1450)$. Finally the vector fields
$\omega^{\mu}$ and $\vec{\rho}^{\;\mu}$ are assigned to the $\omega(782)$ and
$\rho(770)$, respectively and the axialvector fields $f_{1}^{\mu}$ and
$\vec{a}_{1}^{\;\mu}$ to the $f_{1}(1285)$ and $a_{1}(1260)$, respectively
\cite{PDG}. Note, the mass of $a_{1}(1260)$ given by PDG is only an estimate.
According to the Ref.\ \cite{UBW} we fixed the mass of $m_{a_{1}}$ to $1050$
MeV. After shifting the scalar-isoscalar fields $\sigma=(\bar{u}u+\bar{d}d)/\sqrt{2}$ and $G=gg$
by their vacuum expectation values, $\sigma\rightarrow\sigma+\phi$ and
$G\rightarrow G+G_{0}$, a bilinear mixing term $\sim\sigma G$ in
(\ref{Lagrangian}) occurs. This required a diagonalization of the
corresponding effective Lagrangian (realized by a $SO(2)$ rotation) in order
to obtain the physical fields $\sigma^{\prime}$ and $G^{\prime}$:
\begin{equation}
\left(
\begin{array}
[c]{c}%
\sigma^{\prime}\\
G^{\prime}%
\end{array}
\right)  =\left(
\begin{array}
[c]{cc}%
\cos\theta & \sin\theta\\
-\sin\theta & \cos\theta
\end{array}
\right)  \left(
\begin{array}
[c]{c}%
\sigma\\
G
\end{array}
\right)  \text{ ,}%
\end{equation}
where $\theta=\frac{1}{2}\arctan\left[  -4\,\frac{\phi}{G_{0}}\,\frac{m_{0}^{2}%
}{M_{G}^{2}-M_{\sigma}^{2}}\right]$
is the quarkonium-glueball mixing angle.

\section{Results and Discussion}

The model in Eq. (\ref{Lagrangian}) contains ten free parameters:
$m_{0}$, $\lambda_{1}$, $\lambda_{2}$, $m_{1}$, $g_{1}$,
$c_1$, $h$, $\tilde{h}=h_{1}+h_{2}+h_{3}$, $m_{G}$, $\Lambda=\sqrt{11}%
\,C^{2}/(2m_{G})$, where $C$ represents the gluon condensate. Once we used the
masses of $\pi$, $\rho(770)$, $m_{\eta_{N}},$ and $m_{a_{1}}$ as well as the
pion decay constant $f_{\pi}$, we are left with four free parameters:
$C,$ $m_{1}$, $M_{\sigma}$ and $m_{G}$. We obtained them by a $\chi^{2}$
analysis using the five experimental quantities of Table \ref{Table1}
(details in Ref \cite{dilaton}).

\subsection{Scenario with $G^{\prime}$ $\equiv f_{0}(1500)$ and $\sigma
^{\prime}$ $\equiv f_{0}(1370)$.}

Our best fit is obtained for the assignment $\{\sigma^{\prime
},G^{\prime}\}=\{f_{0}(1370),$ $f_{0}(1500)\}.$ We used as input for the $\chi
^{2}$ analysis the following quantities according to \cite{PDG}: masses of the
resonances $f_{0}(1500)$ and $f_{0}(1370)$ (for which we used the mean value
$M_{\sigma^{\prime}}^{ex}=(1350\pm150)$ MeV due to the wide mass range of this
resonance) and the three well-known decay widths of $f_{0}(1500)$:
$\Gamma_{f_{0}(1500)\rightarrow\pi\pi}$, $\Gamma_{f_{0}(1500)\rightarrow
\eta\eta}$ and $\Gamma_{f_{0}(1500)\rightarrow K\bar{K}}$ (see Table
\ref{Table1}).%

\begin{table}[h] \centering
\begin{tabular}
[c]{|c|c|c|}\hline
Quantity & Fit [MeV] & Experiment [MeV]\\\hline
$M_{\sigma^{\prime}}$ & $1191\pm26$ & $1200$-$1500$\\\hline
$M_{G^{\prime}}$ & $1505\pm6$ & $1505\pm6$\\\hline
$G^{\prime}\rightarrow\pi\pi$ & $38\pm5$ & $38.04\pm4.95$\\\hline
$G^{\prime}\rightarrow\eta\eta$ & $5.3\pm1.3$ & $5.56\pm1.34$\\\hline
$G^{\prime}\rightarrow K\bar{K}$ & $9.3\pm1.7$ & $9.37\pm1.69$\\\hline
\end{tabular}%
\caption{Fit in the scenario \{$\sigma
'$, $G'$\} = \{$f_0(1370)$, $f_0(1500)$\}.\label{Table1}}%
\end{table}%

The value of the quarkonium-glueball mixing angle is $\theta=\left(
29.7\pm3.6\right)
{{}^\circ}%
$. This implies that the resonance $f_{0}(1500)$ consists to $76\%$ of a
glueball and to the remaining $24\%$ of a quark-antiquark state. In the case
of $f_{0}(1370)$ we obtain an inverted situation. An important outcome of our
fit is the value of the gluon condensate, $C=(699\pm40)$ MeV, which is in
agreement with lattice QCD results \cite{Lattice}. Note that the gluon
condensate is a essential quantity of QCD and we obtained its numerical
value by use of experimental data. Further consequences and predictions
are given in Table \ref{Table2}.
The decay of $f_{0}(1500)$ into $4\pi$ through the intermediate state of
$\rho\rho$ mesons is calculated by using the $\rho$ spectral function. Our
result is about half of the experimental one. We expect that the intermediate
state consisting of two $f_{0}(500)$ resonances also contributes in this decay
channel, but this resonance is not yet included in the model. The decays of
the resonance $f_{0}(1370)$ are in agreement with the experimental data
regarding the full decay width: $\Gamma_{f_{0}(1370)}=(200$ -- $500)$\ MeV
\cite{PDG}, where our result is around $\Gamma_{\sigma^{\prime}}\simeq
360$\ MeV. Note that the inclusion of the (axial)vector d.o.f. was crucial in
order to obtain the presented results \cite{Paper1,Zakopane}. The artificial
decoupling of (axial)vector states would generate a by far too wide
$f_{0}(1370)$ state. For this reason the glueball-quarkonium mixing scenario
above $1$ GeV has been previously studied only in phenomenological models with
flavour symmetry \cite{Close,longglueball} but not in the context of chirally
invariant models. In Ref. \cite{dilaton} we also investigated the scenario
$\{\sigma^{\prime},G^{\prime}\}=$ $\{f_{0}(1370),$ $f_{0}(1710)\}$ and we have
found that it is not favoured by experimental data.%
\begin{table}[h] \centering
\begin{tabular}
[c]{|c|c|c|}\hline
Quantity & Fit [MeV] & Experiment [MeV]\\\hline
$G^{\prime}\rightarrow\rho\rho\rightarrow4\pi$ & $30$ & $54.0\pm7.1$\\\hline
$G^{\prime}\rightarrow\eta\eta^{\prime}$ & $0.6$ & $2.1\pm1.0$\\\hline
$\sigma^{\prime}\rightarrow\pi\pi$ & $284\pm43$ & -\\\hline
$\sigma^{\prime}\rightarrow\eta\eta$ & $72\pm6$ & -\\\hline
$\sigma^{\prime}\rightarrow K\bar{K}$ & $4.6\pm2.1$ & -\\\hline
$\sigma^{\prime}\rightarrow\rho\rho\rightarrow4\pi$ & $0.09$ & -\\\hline
\end{tabular}%
\caption{Further results regarding the $\sigma' \equiv
f_0(1370)$ and $G' \equiv f_0(1500)$ decays.\label{Table2}}%
\end{table}%

\subsection{Scenarios with $\sigma^{\prime}\equiv f_{0}(500)$}

We have tested the assignments $\{\sigma^{\prime},G^{\prime}\}=$ $\{f_{0}(500),$
$f_{0}(1500)\}$ and $\{\sigma^{\prime},G^{\prime}\}=$ $\{f_{0}(500),$
$f_{0}(1710)\}$ using the new available experimental data of the $f_{0}(500)$
resonance \cite{PDG}. We used for the calculation the mean value of its mass,
$M_{\sigma^{\prime}}^{ex}=(475\pm75)$ MeV. In both assignments the mixing
angle turns out to be small ($\lesssim13^{\circ}$) and this implies that the
state $f_{0}(500)$ is almost a pure quarkonium. The problem of these scenarios
is that the decay into two pions is too narrow, $\Gamma_{\sigma^{\prime
}\rightarrow\pi\pi}\lesssim180$ MeV (as already found in Ref. \cite{dilaton}),
in comparison to the experimental one,
$\Gamma_{f_{0}(500)\rightarrow\pi\pi}(400-700)$ MeV. We thus confirm our result
in Ref. \cite{dilaton} that the scenarios with the resonance $f_{0}(500)$ as a
quarkonium state are not favoured.

\section{Conclusions and Outlook}

We have used a chiral linear sigma model with (axial)vector mesons
and a scalar glueball to study the phenomenology of the scalar-isoscalar
states below $2$ GeV. The best agreement with the present experimental data is
reached when the resonance $f_{0}(1500)$ is predominantly identified with a
glueball state, $\left\vert G\right\rangle
\equiv gg$ and $f_{0}(1370)$ with a quark-antiquark state,
$\left\vert \bar{n}n\right\rangle \equiv\left(  \bar{u}u+\bar{d}d\right)
/\sqrt{2}$. Scenarios in which $f_{0}(500)$ is predominantly a quark-antiquark
state show discrepancies with the experiment.\newline
Ongoing works are the full inclusion of strangeness,\emph{ }$N_{f}=3$
\cite{dick} and eventually the inclusion of a nonet of tetraquarks \cite{mixtetra}.
This may enable us to describe a general mixing scenario of all five scalar-isoscalar
states below the mass of $2$ GeV listed by PDG \cite{PDG}

\section*{Acknowledgments}

The author thanks D. Parganlija, F. Giacosa and D. Rischke for cooperation and
useful discussions and H-QM and HGS-HIRe for funding.

\end{document}